# Superconducting H-mode Structures for Medium Energy Beams

R. Eichhorn and U. Ratzinger

GSI, Planckstr. 1, 64291 Darmstadt, Germany and

IAP, Goethe Universität Frankfurt, Robert-Mayer-Str. 2-4, 60325 Frankfurt, Germany

*Abstract*

Room temperature IH-type drift tube structures are used at different places now for the acceleration of ions with mass over charge ratios up to 65 and velocities between 0.016 c and 0.1 c. These structures have a high shunt impedance and allow the acceleration of very intense beams at high accelerating gradients. The overall power consumption of room temperature IH-mode structures is comparable with superconducting (sc) structures up to 2 MeV/u. With the KONUS [1] beam dynamics, the required transversal focusing elements, e.g. quadrupole triplets can be placed outside of multicell cavities, which is favourable for building sc H-mode cavities. The design principles and consequences to the geometry compared to room temperature (rt) cavities will be described. The results gained from numerical simulations show that a sc multi-gap $H_{21(0)}$-mode cavity (CH-type) can be an alternative to the sc spoke-type or reentrant cavity structures up to beam energies around 150 MeV/u. The main cavity parameters and possible fabrication options will be discussed.

## 1 INTRODUCTION

Linacs based on rt H-mode cavities (RFQ and drift tube structures) are used today in the velocity range from $\beta=0.002$ up to $\beta=0.1$. RF power tests show the capability of IH-cavities to stand about 25 MV/m on-axis field. Beside these high accelerating gradients H-mode cavities allow the acceleration of intense beams [2]. One aspect of the investigations started at GSI and IAP is to extend the velocity range of the H-mode cavities up to $\beta=0.5$ by using the $H_{21(0)}$ or CH-mode.

Many future projects (the Accelerator Driven Transmutation Project ADTP[3], the European Spallation Source ESS[4] or the Heavy Ion Inertial Fusion HIIF [5]) are based on the availability of efficient accelerating cavities with properties like mentioned above, which additionally could be operated in cw mode. It is commonly accepted that above an energy of 200 MeV/u superconducting cavities are superior to rt structures. By combining the advantages of CH-mode cavities with the benefits of superconductivity, effective ion acceleration at high duty cycle will be possible. For high current proton beams the injection energy will be around 10 MeV, while for heavy ions the injection energy may become as low as 1 MeV/u. The CH-structure is efficient for beam energies up to 150 MeV/u.

This paper describes the properties of H-mode cavities and especially the CH-type. Important is the application of the KONUS beam dynamics [1], resulting in long, lens free accelerating sections housed in individual cavities. This opens the superconducting option, as the magnetic field of cavity internal quadrupoles cannot be easily shielded well enough to avoid frozen current contributions.

The results from numerical simulations of three $H_{21(0)}$-type cavities with different resonant frequencies and velocity profiles will be reported and first mechanical construction layouts will be presented.

## 2 H-MODE CAVITIES

The IH-DTL (Interdigital H-mode or $H_{11(0)}$) has become a standard solution for heavy ion acceleration. The CERN Pb-injector installed in 1994 [6] is one example. The new high current injector at GSI [2,7] consists of the first IH-DTL designed for very heavy ions with A/q = 65 including space charge effects of considerable strength at the design current of I/emA = 0.25·A/q. An important property of H-type DTLs is their high acceleration efficiency, i.e. they provide a high shunt impedance, especially at low β-values up to 0.2.

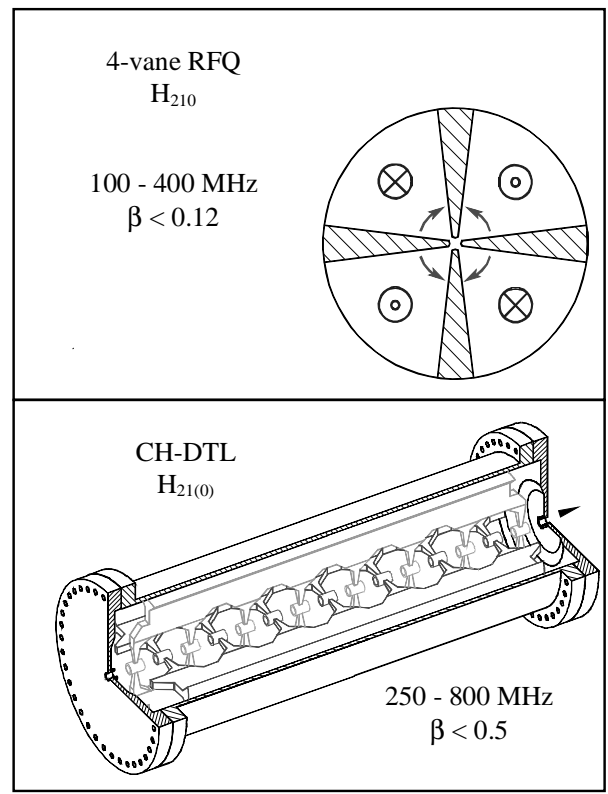

Figure 1: The CH-mode structure family: The main direction of the magnetic RF field is oriented parallel and anti parallel with respect to the beam axis.

The $H_{210}$-mode is already used in accelerator physics: the 4 vane-RFQ is well established for proton and light ion acceleration [8]. A competitive H-mode drift tube cavity in the velocity range from β = 0.05 to 0.5 may become the $H_{21(0)}$-mode CH structure, which can be deduced from the 4-vane RFQ by cutting down the vanes around the aperture and replacing the electrodes by drift tubes. These drift tubes are connected by two stems with the girders of identical RF potential (see fig. 1). The dipole mode that usually causes trouble during the tuning procedure of the 4-vane RFQ is short-circuited by the stems along the whole cavity. The analytically estimated shunt impedance assuming slim drift tubes without quadrupoles is about a factor 1.4 higher compared to the IH-cavity. Numerical simulations have shown that CH-cavities for resonance frequencies up to around 800 MHz can be realised. This allows to close the velocity gap between RFQs and Coupled Cavity Linacs (CCL).

## 3 DESIGN OF A SUPERCONDUCTING CH-CAVITY

The CH-cavity exceeds by far the mechanical rigidity of IH-tanks. This opens the possibility to develop superconducting multi-cell cavities [9]. So far only 2- and 3-cell sc structures were realised for low beam velocities.

The different prototype cavities discussed are modules of an accelerator design investigated for high current proton acceleration. The parameters of these prototype cavities are given in tab.1.

The RF behavior of the resonators was studied with an analytical model [10] that allows a first optimization step of the fundamental cavity parameters. The consequent numerical simulations of the resonators were done using the MAFIA package [11]. All parameters predicted with the analytical model were confirmed by MAFIA within ±10 % deviation.

As a first step of the design process, only two accelerating gaps have been computed. Starting with a rt design the drift tubes and the shape of the stems were optimized. Special care was taken on the magnetic surface field trying to keep it well below the BCS-Limit of Niobium. By increasing the stem cross section, the magnetic surface field could be reduced by aprox. a factor of two, reaching up to 30 mT in the actual design.

Table 1: Parameters and expected performance of the prototype cavities

| frequency (MHz) | 352 | 433 | 700 |
|---|---|---|---|
| particles | protons | | |
| injection energy (MeV) | 10.9 | 10.9 | 130 |
| eff. voltage gain (MV/m) | 6.7 | | |
| particle velocity (v/c) | 0.17 | 0.17 | 0.5 |
| mode | $H_{21(0)}$ (CH) | | |
| gap number | 18 | 18 | 10 |
| length of the cavity (m) | 1.43 | 1.01 | 1.07 |
| drift tube aperture (mm) | 25 | 25 | 10 |
| transit time factor | 0.8-0.85 | | |
| tank radius (mm) | 185 | 130 | 105 |
| stored energy (J) | 12.6 | 4.1 | 4.3 |
| $E_{max}/E_{acc}$ | 4.1 | 3.9 | 5.2 |
| $B_{max}/E_{acc}$ (mT/(MV/m)) | 5.8 | 3.7 | 8.8 |
| $R/Q_0$ (kΩ/m) | 2.7 | 4.7 | 2.6 |
| geometric factor (Ω) | 163 | 160 | 208 |
| Q-factor (4K,Nb) | $4.3·10^9$ | $2.7·10^9$ | $1.4·10^9$ |
| diss. power (4K,Nb) (W) | 6.5 | 4.0 | 13.8 |
| shunt impedance (MΩ/m) | 44 | 68 | 39 |
| Q-factor (rt, copper) | 16400 | 14600 | 14800 |

This optimization process caused an increase in the capacitive load and thus lowered the shunt impedance by 20 %, which is no drawback in case of a sc cavity. The maximum electric field (27.5 MV/m) has been found at the drift tube ends.

After this optimization step the whole cavity was computed to study the impacts on the field flatness. As the magnetic flux bends from one to the neighboring sector at its ends (see also fig. 1) one expects high magnetic surface fields in this region. The girder undercuts, used in rt IH-mode and 4-vane cavities successfully to create the zero mode are not the right choice for a sc cavity. Therefore a careful redesign had to be performed. The lowest surface fields were yielded by combining two modifications: The tank radius is increased by 17 % between the end flange and the first drift tube stem. To assure the field flatness, i.e. a constant accelerating field along the whole cavity, the local capacity at the cavity ends had to be further increased. This was attained by forming thicker stems. As a measure to reduce the surface currents at the cavity end flanges, which may be sealed by an indium joint, half drift tubes were introduced.

Figure 2 shows the layout of the cavity. Table 1 summarizes the results of the simulations. Comparing these parameters to that of existing cavities [12] displays the potential of CH-cavities.

## 4 FABRICATION OPTIONS

Up to now only two different options in fabricating the cavity are considered:

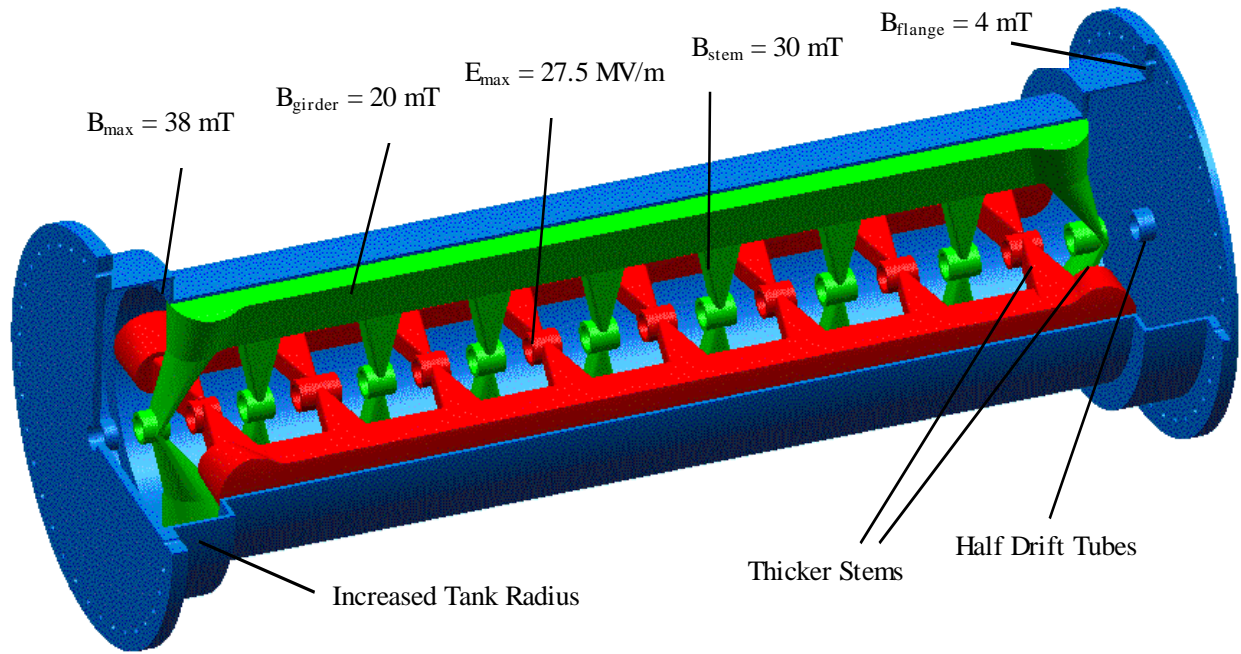

Figure 2: Three dimensional view of a 352 MHz CH mode cavity for β = 0.17. The cavity dimensions can be found in tab. 1. Also shown are some field values calculated with MAFIA, especially the maximum values of the magnetic (38 mT at both resonator ends) and of the electric field (27.5 MV/m on the drift tube surface).

- Like it is done with the normal conducting IH-mode resonators, one fabricates the drift tubes and the stems out of bulk copper mounted inside a copper plated steel tank. The deposition of the superconducting surface layer is done by sputtering with niobium. Though the behavior of thin niobium films is well known and meet the requirements, it might need much effort for sputtering this complex geometry with the needed precision. The progress in this technology made elsewhere [13] shows however the great potential of this option.
- On the other hand one can think of building the cavity out of bulk niobium components. In that case there exists a lot of experience in electron beam welding. The first superconducting RFQ, which has currently past the cold low level tests successfully, has been built this way [14].

The third possibility, namely leadplating a copper resonator has been abandoned because of the limited performance of cavities build this way, as for example in Legnaro [15]. Even though the magnetic surface flux of our resonator is below the BCS-limit for lead, electric surface fields of up to 28 MV/m seems to be beyond the limit of that technology.

## 5 OUTLOOK

The normal conducting H-mode cavities showed their suitability for ion acceleration at low energies in many accelerator laboratories. Our investigations indicate that CH-mode cavities are well suited to design superconducting resonators. The results of the numerical simulation of such a cavity are very promising. Using state of the art technology the fabrication of a superconducting CH-mode cavity should be possible [13,14]. The design, construction and rf test of an 352 MHz prototype cavity are scheduled at IAP, Frankfurt University.

## REFERENCES

[1] U. Ratzinger, Nucl. Intr. Meth., A 415 (1998) 229.
[2] U. Ratzinger, Proc. Int. Linac. Conf., Geneva, CERN 96-07 (1996) 288.
[3] R.A. Jameson et al., Proc. Europ. Part. Acc. Conf., Berlin, (1992) 230.
[4] ESS-CDR, eds.: I. Gardener, H. Lengeler and G. Rees, ESS-Report 95-30-M (1995).
[5] U. Ratzinger, Nucl. Intr. Meth., in press.
[6] H.D. Haseroth, Proc. Int. Linac Conf., Geneva, CERN 96-07 (1996) 283.
[7] W. Barth et al., this conference
[8] K.R. Crandall et al, Proc. Int. Linac Conf., Montauk, BNL51134 (1979) 205.
[9] R. Eichhorn and U. Ratzinger, Proc. 9th RF Supercond. Workshop, Santa Fe (1999) in press.
[10] U. Ratzinger, CAS 2000 Seeheim (2000) submitted
[11] T. Weiland, *MAFIA - A Three Dimensional Electromagnetic CAD System for Magnets, RF Structures and Transient Wake Field Calculations* (1998).
[12] K. Shepard, Nuc. Instr. Meth. A 382 (1996) 125.
[13] V. Palmieri et al., LNL-INFN (REP) 149/99.
[14] G. Bisoffi et. al., Proc. Europ. Part. Acc. Conf., Vienna ,(2000) in press.
[15] A.-M. Porcelatto et al., Proc. 9th RF Supercond. Workshop, Santa Fe (1999) in press.